\documentclass[12pt,preprint]{aastex}
\usepackage{apjfonts}

\def\ltsima{$\; \buildrel < \over \sim \;$}
\def\simlt{\lower.5ex\hbox{\ltsima}}
\def\gtsima{$\; \buildrel > \over \sim \;$}
\def\simgt{\lower.5ex\hbox{\gtsima}}

\slugcomment{}

\shorttitle{SGR 1806-20 and Variability Limit} \shortauthors{Vietri et al.}

\begin{document}

\title{SGR1806-20: evidence for a superstrong Magnetic Field from Quasi Periodic Oscillations}

\author{Mario Vietri}
\affil{Scuola Normale Superiore, 56100 Pisa, Italy }
\email{m.vietri@sns.it}

\and

\author{Luigi Stella and Gian Luca Israel}
\affil{INAF - Osservatorio Astronomico di Roma, \\ Via Frascati 33,
  00040 Monteporzio Catone, Italy.}
\email{stella@mporzio.astro.it} \email{gianluca@mporzio.astro.it}

\begin{abstract}

Fast Quasi-Periodic Oscillations (QPOs, frequencies of $\sim 20 -
1840$~Hz) have been recently discovered in the ringing tail of giant
flares from Soft Gamma Repeaters (SGRs), when the luminosity was of
order $10^{41}-10^{41.5}$~erg/s. These oscillations persisted for
many tens of seconds, remained coherent for up to hundreds of cycles
and were observed over a wide range of rotational phases of the
neutron stars  believed to host SGRs. Therefore these QPOs must have
originated from a compact, virtually non-expanding region inside the
star's magnetosphere, emitting with a very moderate degree of
beaming (if at all). The fastest QPOs imply a luminosity variation
of $\Delta L/\Delta t \simeq 6 \times 10^{43}$~erg~s$^{-2}$, the
largest luminosity variation ever observed from a compact source. It
exceeds by over an order of magnitude the usual Cavallo-Fabian-Rees
(CFR) luminosity variability limit for a matter-to-radiation
conversion efficiency of 100\%. We show that such an extreme
variability can be reconciled with the CFR limit if the emitting
region is immersed in a magnetic field $\gtrsim 10^{15}$~G at the
star surface, providing independent evidence for the superstrong
magnetic fields of magnetars.

\end{abstract}


\section{Introduction}

Soft Gamma Repeaters are a small class of galactic sources of X and
soft gamma radiation. They have spin periods of $\sim 5\div10$~s,
display a secular spin-down with timescales of $\sim
10^4\div10^5$~yr and do not possess a companion.
Unlike radio pulsars, the rotational energy
loss of SGRs is a factor of $10\div100$ too small to explain
their persistent emission, typically $\sim 10^{33}\div10^{34}$ erg/s
(see e.g. \citet{WooTho06}). Like Anomalous X-ray Pulsar (AXPs,
\citet{MeSte95}), with whom they share a number of properties, SGRs
are believed to host magnetars, neutron stars the emission of which
is powered by the decay of their superstrong (internal)
magnetic field ($B> 10^{15}$~G, \citet{DT92,TD93}).

The name-defining characteristic of SGRs is that they show periods
of activity in which recurrent short bursts are emitted, with peak
luminosities of $\sim 10^{38}\div10^{41}$~ergs~s$^{-1}$ and
sub-second durations. The characteristics of the three giant flares
observed so far in about 30~yr of monitoring are much more extreme.
Their initial, tenths-of-seconds-long spike releases enormous
amounts of energy, $\sim 10^{44}$~ergs in the 1979 March 5 event
from SGR 0526-66 \citep{Maz79} and the 1998 August 27 event from SGR
1900+14 \citep{Hur99,Fer99} and as much as $ 5 \times 10^{46}$~erg
in the 2004 December 27 event from SGR1806-20
\citep{Hur05,Mere05,Palm05,Teras05}. After the initial spike, giant
flares display a very bright ringing tail lasting hundreds of
seconds and releasing a total energy of about $\sim 10^{44}$~ergs.
The emitted spectrum is roughly thermal, with a blackbody equivalent
temperature of $\sim 5$~keV in the case of SGR1806-20 \citet{Hur05}.

The highly super-Eddington luminosities of the recurrent bursts of
SGRs and especially of their giant flares make models involving
accretion energy not viable.
According to the magnetar model, the emission of SGRs (and AXPs)
draws from their extremely high magnetic fields
\citep{TD95,TD96,TD01}. Within this model the neutron star interior
is characterized by a wound-up, mainly toroidal magnetic field
configuration with $B_t>10^{15}$~G. A less intense (mainly poloidal)
field emerges out of the star magnetosphere, causing spin-down via
rotating dipole losses at the observed rate \citep{TD93,TD01}
(dipole B-field strength of $B_d \sim 7.8\times 10^{14}$~G). Energy
propagates to the neutron star magnetosphere through Alfv\'{e}n
waves driven by local ``crust-quakes" and giving rise to recurrent
bursts with a large range of amplitudes. Large-scale rearrangements
of the internal magnetic field or catastrophic instabilities in the
magnetosphere are invoked to explain the sudden release of very
large amounts of energy that occurs in giant flares
\citep{TD01,Lyu03}. A fireball of plasma expanding at relativistic
speeds breaks out of the star's magnetosphere, causing the initial
sub-second spike of giant flares. The ringing tail that follows
results most likely from the part of the fireball that remains
trapped in the star's magnetosphere. The energy release in the
ringing tail yields a limit for the external field of magnetars
($\simgt 10^{14}$~G) in agreement with the values inferred from
spin-down dipole losses \citep{TD95,TD01}, while an analysis of the
initial time scales gives evidence in favor of the crustal cracking
mechanism \citep{Sch05}.

To confirm this model it is thus essential to measure the surface
magnetic field, which complements the measurement of the dipole
component (\citet{Woo02}), to include higher order multipoles. In
this paper we present an interpretation of the QPOs observed in SGR
1806-20 which provides a lower limit ($B \gg B_q \approx 4.4\times
10^{13}\; G$) to the {\it surface} field. The argument is based upon
a seldom-used constraint (\citep{CaRe78,FaRe79,Fab79}) that puts a
very strong upper limit on the time-scale on which significant
luminosity variations can take place. In the next section we discuss
the relevant observations; in Section 3 we will re-derive the limit
and show that it is largely violated by QPOs in SGR 1806-20; after
vainly trying to circumvent it, we will show that it can be
reconciled with observations only if $B\gg B_q$.

\section{Quasi Periodic Oscillations in Giant Flares of SGRs}

Recent studies led to the discovery that the X-ray flux of the
ringing tail of SGRs' giant flares is characterized by fast Quasi
Periodic Oscillations, QPOs \citep{Isr05}. Different QPO modes were
detected, some of which were excited simultaneously. The ringing
tail of the December 2004 event from SGR~1806-20 displayed clear QPO
signals at about 18, 30, 93, 150, 625 and 1840~Hz (\citet{WaSt06}).
Similarly, QPOs around frequencies of 28, 54, 84 and 155 Hz were
detected during the ringing tail of the 1998 giant flare of
SGR~1900+14 \citep{StWa05}, while hints for a signal at $\sim$43 Hz
were found in the March 1979 event from SGR 0526-66 \citep{Bar83}.
These QPOs show large variations of the amplitude with time and,
especially, of the phase of the spin modulation in the giant flare's
tail.

The similarity in some of the QPO modes and frequencies across
different SGRs suggests that the production mechanism is the same. A
likely interpretation involves the excitation of neutron star
oscillation modes, whose expected eigenfrequencies match some of the
observed QPOs peaks (\citep{Dun98,Isr05,Pir05}). Infact, if giant
flares result from large scale fracturing of the crust induced by
instabilities of the internal magnetic fields, then the excitation
of crustal and (possibly) global neutron star modes is to be
expected (\citep{Lev06}). Regardless of the exact mechanism driving
these oscillations we are concerned here with the extremely large
and fast luminosity variations of the QPOs.

We concentrate on the signals with the largest luminosity time
derivative, i.e. the 625 and 1840~Hz QPOs from SGR 1806-20
(\citet{WaSt06}). The power spectrum peaks through which these QPOs
are revealed, are a few Hz wide, testifying that their signal
remained coherent for hundreds of cycles. The signal shape must be
close to sinusoidal, as evidenced by the absence of detectable
harmonic signals. The 625 and 1840~Hz QPOs were detected only in a
$\sim 50$~s long interval of the ringing tail, about 200~s after the
initial spike, and were especially prominent over a $\sim 140$~deg
interval in rotational phase. The QPO amplitude reached a maximum in
this phase interval over two consecutive rotation cycles: for both
signals the rms amplitude was $a_{rms}\sim 18$\%. Approximating the
QPOs with sinusoids, we estimate their highest luminosity derivative
as $\Delta L/\Delta t = 2^{3/2} \pi L a_{rms} \nu_{QPO}$, with
$\nu_{QPO}$ the QPO frequency. Here $L \sim 10^{41}$~ergs~s$^{-1}$
is the luminosity in the relevant section of the ringing tail (for
the likely source distance of 15 ~kpc). This gives $\Delta L/\Delta
t \sim 1\times 10^{44}$ and $ 3\times 10^{44}$~ergs~s$^{-2}$ for the
625 and 1840~Hz QPOs, respectively. The effects of beaming might
decrease these values somewhat, but not by a very large factor. In
fact, these QPO signals were observed over a large interval of
rotational phases (about $\sim 140$~deg), translating into
approximately the same azimuthal range of emission angles.  It is
natural to assume a comparably large angular spread in latitude
(unless the neutron star rotation axis is very close to our line of
sight, which is unlikely given the large amplitude of the rotation
modulation and the size of the emission region, see below). This
gives a solid angle of order $\sim \pi$~ster. Adopting this beaming
factor the luminosity derivatives above reduce to $\sim 2\times
10^{43}$ and $ 6\times 10^{43}$~ergs~s$^{-2}$; this are the values
that we adopt in the following discussion.

We stress here, because this is essential to our argument (to be
presented shortly), that, together with the QPOs, in the ringing
tail a strong modulation at the star's spin period is clearly
present, similar in relative amplitude and shape to the modulation
observed when the source is in its quiescent state: this (together
with the lack of significant amounts of beaming) indicates that the
emission in the ringing tail originates from a region that remains
stably anchored to the star's magnetosphere, and thus that
relativistic bulk motions are not present at this stage of the flare.

The blackbody temperature and luminosity in the ringing tail
translates into a lower limit on the size of the emitting region of
about $\sim 30$~km, {\it i.e.} substantially larger than the neutron
star. On the other hand the black body-like spectral shape testifies
that the emitting region is optically thick (or at least effectively
thick), implying a scattering optical depth $\gg 1$. We remark that
the size estimate, $\approx 30\; km$, will play an important role in
the following.

\section{The Cavallo-Fabian-Rees Variability Limit}

There is a well--known limit on the rate of change of the luminosity
of any given source, which we briefly summarize here
(\citet{CaRe78,FaRe79,Fab79}, see also \citet{Law80,Hos89}). Suppose
a source undergoes a large luminosity variation on a time-scale
$\Delta t$, and there is a luminosity variation, over this
time-scale, $\Delta L$. The total energy released within $\Delta t$
is related to the total mass within the source dimension $R$ by
\begin{equation}\label{energycontent}
\Delta L \Delta t = \eta \frac{4\pi}{3}R^3 n m_p c^2\;,
\end{equation}
where $n$ is the average baryon density, and $\eta$ is the energy
extraction efficiency. The time-scale $\Delta t$ must
obviously exceed the time over which the photons manage to diffuse
out of the source:
\begin{equation}\label{timevar}
\Delta t > \frac{R}{c} (1+\tau_T)\;
\end{equation}
where $\tau_T = \sigma_T n R$ is the Thomson optical depth
and $\sigma_T$ the Thomson cross-section. Eliminating $R$
from the first two equations, we find
\begin{equation}
\Delta t > \frac{3}{8\pi}\frac{\sigma_T}{m_p c^4} \frac{\Delta
L}{\eta}\frac{(\tau_T+1)^2}{\tau_T}\;.
\end{equation}
Regarded as a function of $\tau_T$, the above has a minimum for
$\tau_T = 1$, which yields the limit
\begin{equation}\label{cavallo}
\Delta t > \frac{3}{2\pi}\frac{\sigma_T}{m_p c^4} \frac{\Delta
L}{\eta}\;.
\end{equation}{}
The above limit is remarkable in that it is independent of both $R$
and $n$, or any combination thereof: only the dependence on $\Delta
L$, a directly observable quantity, is left. The
Cavallo-Fabian-Rees, CFR, limit thus writes
\begin{equation}\label{CFR}
\Delta L /\Delta t < \eta\ 2\times 10^{42}erg\; s^{-2}\;.
\end{equation}{}

The 625 and 1840~Hz QPO signals from SGR1806-20 exceed the CFR limit
by about an order of magnitude\footnote{We note that also the slower
$\sim 90-150~Hz$ QPOs signals in the giant flares' tail of
SGR1806-20 and SGR1900+14, exceed the CFR limit, though by a smaller
factor.}: the largest value found in the previous section is $\Delta
L/\Delta t = 6\times 10^{43}\; erg\; s^{-2}$, which is a whole
factor $30/\eta$ larger than the CFR's limit.

In order to appreciate how hard it is to circumvent this limit,
notice the {\it in situ} re-acceleration of electrons does not help,
because it does not change the energies reached by protons; if
protons were to escape, leaving electrons behind to be
re-accelerated at will, Coulomb forces would quickly make the escape
of protons impossible. Nor will having relativistic protons help, as
one might think, since this would imply energies per electron to be
radiated $\approx \gamma m_p c^2\gg \eta m_p c^2$, because the
energy transfer from protons to electrons is too slow. To show this,
let us consider what happens when protons transfer promptly to the
electrons their energy gain: protons may then be relativistic, in
which case the maximum energy which can be extracted from each of
them is $\gamma m_p c^2$, rather than the more sedate $\eta m_p
c^2$. This however appears like an unlikely way out, because, even
admitting that the electrons radiate very promptly their internal
energy, the time-scale on which protons manage to transfer to the
electrons their internal energy is much longer than the above limit.
To see this, we idealize the situation as one where electrons are
cold ({\it i.e.} Newtonian) as a result of their short cooling
timescales, while the protons are still relativistic. The energy
transfer rate is given by the usual formula
\begin{equation}
-\frac{dE_p}{dx} = \frac{2\pi e^4 n_e}{m_e v^2}\left( \ln(\frac{2m_e
v^2 W}{\hbar^2\omega_p^2})+1-2\frac{v^2}{c^2} \right)\;,
\end{equation}
where $n_e$ is the electron number density, $\omega_p = \sqrt{4\pi
e^2 n_e /m_e}$ the plasma frequency, $v \approx c$ the proton speed,
and $W \approx 2E_p^2/(m_e c^2)$ is the maximum energy transfer in
the not too extreme limit, $\gamma_p \lesssim m_p/m_e$ (but
considering the opposite limit, $\gamma \gtrsim m_p/m_e$ would
change the argument very little since $W$ only appears as an
argument to a logarithm).

The energy transfer timescale is of course $t_e \equiv E_p/(c\
dE_p/dx)$, which, for this whole idea to work, must be shorter than
$\Delta t$. The condition $t_E < \Delta t$ can also be rewritten as,
in the limit $v^2 \approx c^2$:
\begin{equation}
\frac{\gamma}{\ln(\frac{2m_e v^2 W}{\hbar^2\omega_p^2})-1} <
\frac{3\sigma_T}{m_p c^4}\frac{L}{\eta} \frac{e^4 n_e}{m_p m_e
c^3}\;.
\end{equation}
Inserting the numerical values for the luminosity of SGR 1806-20
during the ringing tail and for the typical magnetospheric density
for a pulsar with a {\it normal} magnetic field we find
\begin{equation}
\gamma \lesssim 10^{-8} \frac{1}{\eta} \frac{L}{3\times 10^{41}\;
erg s^{-1}} \frac{n_e}{10^{10}\; cm^{-3}}\;.
\end{equation}
In order to reconcile the CFR limit with observations, we need
$\gamma \approx 30$, corresponding to a magnetospheric density
$n\approx 3\times 10^{19}\; cm^{-3}$. While it is certainly true
that plasma outside pulsars need not be charge--separated, still
this density would exceed the minimum ({\it i.e.}, charge separated)
value by more than $9$ orders of magnitude, making it unlikely that
such a plasma may exist.

This argument must be modified in the presence of a pair plasma,
where pair creation processes can easily lead to a strong increase
in $n_e \propto T^3$. Still, we know that astrophysical pair plasmas
are thermally regulated \citep{PiKr95}, so that their temperatures
always lie around $T \lesssim 1\; MeV$. At these temperatures, $n_e
\approx 10^{31}\; cm^{-3}$, but we also know that $E_p \ll m_p c^2$,
so that the limit of eq. \ref{cavallo} still holds, provided the
total energy density is still dominated by the rest mass of the
baryons. When instead the total energy density is dominated by the
pairs (in other words, when little or no baryons
are admixed), eq. \ref{timevar} 
obviously still holds (with $\tau_T$ now indicating the total
optical depth due to the pairs and possibly to photons as well),
while eq. \ref{energycontent} must be rewritten as
\begin{equation}
\Delta L\Delta t = \eta \frac{4\pi}{3} R^3 n m_e c^2
\end{equation}
where we used the fact that $\gamma\approx 2$ for the electrons. The
new limit for the variability is
\begin{equation}\label{cavallo2}{}
\Delta t > \frac{3}{2\pi}\frac{\sigma_T}{m_e c^4} \frac{\Delta
L}{\eta} = 134 \;s \frac{1}{\eta}\frac{\Delta L}{3\times 10^{41}
erg\; s^{-1}}\;,
\end{equation}{}
which is even more stringent than eq. \ref{cavallo}.

The CFR limit might fail if coherent phenomena are involved, but we
are unaware of any coherent mechanism working in the X--ray region
of the spectrum where QPOs are observed, rather than the usual radio
band.

The classic way to circumvent the CFR limit is by means of
relativistic effects. After all, GRBs' light-curves often display
millisecond variability, and even when they don't they turn on on
timescales of a second or less, reaching $L \approx 10^{50}\ erg \;
s^{-1}$: this clearly violates eq. \ref{cavallo} by eight to ten
orders of magnitude (for $\eta =1$). This is due to a combination of
relativistic aberration, blueshift and time contraction, which can
drastically increase the luminosity and decrease the variability
timescale in the observer's frame. Also BL Lac objects and Quasars
have luminosity derivatives that exceed by orders of magnitude the
CFR limit (see e.g. \citep{Bas83}), even after their luminosity is
corrected for beaming. This explanation is  probably correct also
for the giant flare of SGR~1806-20, where values as high as $\Delta
L/\Delta t \sim 10^{47}\; ergs\;s^{-2}$, for isotropic
ejection\footnote{Beaming in the ejection, which is presently
unknown, could decrease this value substantially.} are observed: the
observation of a radio halo expanding at relativistic speed
testifies that the observed variability in the initial spike of the
giant flare is most likely magnified by relativistic effects as
well.

However, this explanation is not suitable for the QPOs in the
ringing tail of the same flare: in fact, the presence of modulations
at the spin period, with similar pulse shape and amplitude to those
during quiescent periods, makes the existence of relativistic bulk
motions very unlikely during this phase. Also, one should remember
that the photosphere size is estimated to be $\approx 30\; km$ at
all times during the ringing tail, which implies that emission
during this phase is due to material stably anchored to the pulsar.
Lastly, the observed temperatures ($T \approx 5\; keV$) testify to
the gas having thermal speeds much below the escape velocity at that
radius. For these three reasons, we deem relativistic effects an
unlikely explanation for the violation of eq. \ref{cavallo}.

The last way in which the CFR limit can fail is the one we propose:
the scattering cross-section may differ from Thomson's because of
the presence of a strong magnetic field $B$, exceeding the quantum
value $B_q = m^2 c^3/(e \hbar) = 4.4\times 10^{13} G$. In this case,
the scattering cross section for the ordinary (O) and extraordinary
(E) modes, and for the conversion of photons into the other state,
are given by \citet{Mes92}, when the dielectric tensor is dominated
by vacuum polarization effects, as:
\begin{eqnarray}\label{meszaros}
d\sigma_{O\rightarrow O} = \frac{3}{4}\sigma_T \sin^2\theta
\sin^2\theta' d\cos\theta' \nonumber\\
d\sigma_{O\rightarrow E} = \frac{3}{8}\sigma_T\left(\frac{\epsilon
B_q}{m_ec^2 B}\right)^2\cos^2\theta d\cos\theta' \nonumber\\
d\sigma_{E\rightarrow O} = \frac{3}{8}\sigma_T\left(\frac{\epsilon
B_q}{m_ec^2 B}\right)^2\cos^2\theta' d\cos\theta' \nonumber\\
d\sigma_{E\rightarrow E} = \frac{3}{8}\sigma_T\left(\frac{\epsilon
B_q}{m_ec^2 B}\right)^2 d\cos\theta'\;,
\end{eqnarray}
where $\epsilon$ is the photon energy, and $\theta$ and $\theta'$
are the angles between the photon momentum before and after the
diffusion, respectively, with the direction of the magnetic field.

This equation shows immediately that, for photons emitted in the
extraordinary mode, the cross section is reduced, with respect to
the Thomson value, by a factor $\approx (\epsilon B_q /(m_e c^2
B))^2$. Thus, to bring the observed value, $\bigtriangleup
L/\bigtriangleup t = 6\times 10^{43}\; erg\; s^{-1}$, in agreement
with eq. \ref{cavallo}, we just need to have $(\epsilon B_q/(m_e c^2
B))^2 \lesssim \eta/30$; here we take for $\epsilon$ the value
$\epsilon \approx 14\; keV$, which is the peak of the Planck
distribution for the observed temperature $T = 5\; keV$. So our
conclusion is that the QPOs' luminosity variation agrees with the
CFR limit provided
\begin{equation}{}
B \gtrsim 1.5 B_q\left(\frac{0.1}{\eta}\right)^{1/2} \approx
6.6\times 10^{13}\; G\;.
\end{equation}
A technical comment is in order at this point: the description of
radiation transfer in terms of separate modal propagation is not
always adequate (\citet{Lai03}), because of mode collapse. However
this effect seems to be mostly relevant for even higher fields ($B
\gtrsim 7\times 10^{13} \; G$) than those derived here, which means
that our naive treatment is probably justified.

We now remark that this lower limit applies to the field close to,
but not {\it at} the surface of SGR 1806-20, because, as discussed
in the previous section, emission from the ringing tail is generated
within $30\; km$. At this distance from the star surface, the dipole
field inferred from pulsar spin-down ($B = 7.8\times 10^{14}\; G$,
\citet{Woo02}), is $2.9\times 10^{13}\; G (R_{ns}/10\; km)^3$, which
is smaller than the limit just derived, as expected if higher order
multipoles are relevant in the star vicinity.

Given the rapid decrease of the dipole field (and {\it a fortiori}
of the other multipoles) with distance from the star, the surface
magnetic field must certainly satisfy
\begin{equation}\label{finallimit}{}
B \gtrsim 1.8\times 10^{15}\; G \left(\frac{10\;
km}{R_{ns}}\right)^3 \left(\frac{0.1}{\eta}\right)^{1/2}\;,
\end{equation}
where a dipole--like radial dependence has been assumed between
$30\; km$ and the star surface, located at $R_{ns}$; we shall
refrain from making more elaborate hypothesis about the structure of
the magnetic field within $R = 30\; km$ (\citet{TD01}), because our
aim is simply to provide a {\it lower limit} to the surface field,
for which this minimum hypothesis (pure dipole) is fully adequate.

\section{Discussion}

Our final result is the lower limit in eq. \ref{finallimit}. This
value is close to that of the dipole field, as inferred from pulsar
spin-down, $B = 7.8\times 10^{14}\; G$, but our estimate includes
higher order multipoles, at least at $30\; km$. It is thus
completely independent of the estimates from pulsar spin-down. We
note that the limit on $B$ derived by requiring that the magnetic
field close to the star can prevent the escape of the "trapped
fireball" includes also the contribution of higher multipole
components (\citet{TD01}), but gives a substantially lower magnetic
field ($B>10^{14}\; G$) than the limit derived above.

There is an easy way to test our interpretation of the failure of
the CFR limit. The O- and E-mode photospheres of the trapped
fireball are located at different heights in the star's
magnetosphere, as a result of the different electron scattering
cross sections, which therefore sample regions of different B-field
strengths and orientations. The polarization fraction and angle of
the emerging X-ray flux should thus be modulated with the phase of
the QPO signal as a result of the varying (relative) intensity of
the O- and E-mode photon component. This in principle gives a clear
test for the correctness of our interpretation, which might perhaps
become verifiable in the future.

Photons in the O-mode suffer a strong Comptonization (\citet{TD01}).
If the atmosphere were due to pure scattering, the ensuing photon
distribution would differ from a blackbody at low photon frequency,
$E_\nu \propto \nu^3$ instead of $\propto \nu^2$, a result due to
photon number conservation. At first sight, one might think that
there are many processes which may lead to photon absorption and
emission in a strong magnetic field, like photon splitting or pair
creation via $\gamma + B$ (resulting then in pair annihilation, and
photon energy downgrading via Compton recoil), which make photon
number conservation unlikely. At the same time, there are important
radiation transfer effects taking place, which obviously tend to
favor flatter spectra at low energies; also, all rates for photon
emission and absorption depart from their values in the absence of
magnetic field. Detailed computations (\citep{Lyu02}) show that the
ensuing spectrum is flatter, not steeper, than a blackbody; and we
remark here that, despite many calibration uncertainties, fits to
the spectrum seem to favor spectra flatter than a blackbody, with a
preference for thermal bremsstrahlung of temperature $T = 30\; keV$.

We have presented our argument by stressing its independence from
the estimates of the dipole field from pulsar spin-down, and from
Thompson and Duncan's (\citet{TD93, TD95, TD96}) model, but it
should be obvious that the limit in eq. \ref{finallimit} is
perfectly consistent with the previous measurements, and the model
itself.

\acknowledgments

This work was partially supported through ASI and MIUR grants.


\end{document}